\documentclass{article}
\usepackage{subcaption}
\usepackage{amsmath}
\usepackage{amssymb}
\usepackage{amsthm}
\usepackage{graphicx} 
\usepackage{fullpage}
\usepackage{color}
\usepackage[backend=biber]{biblatex}
\usepackage{tikz}
\usepackage{float}
\usepackage{authblk} 
\usetikzlibrary{arrows.meta,decorations.markings}

\tikzset{
  flow/.style={
    postaction={decorate},
    decoration={markings,
      mark=at position #1 with {\arrow{Stealth[length=3mm]}}}
  }
}
\newtheorem{prop}{Proposition}
\newtheorem{lemma}[prop]{Lemma}
\newtheorem{remark}[prop]{Remark}
\newtheorem{cor}[prop]{Corollary}

\title{A minimal model for rate-induced tipping to extinction}
\date{June 2026}

\begin{document}
\author[$1,2$]{Brian Hennessy} 
\author[$1$]{Nikola Popovi\'c}
\author[$1$]{Zak Sattar \thanks{z.sattar@sms.ed.ac.uk}}
\affil[$1$]{School of Mathematics and Maxwell Institute for Mathematical Sciences, University of Edinburgh, James Clerk Maxwell Building, King’s Buildings, Peter Guthrie Tait Road, Edinburgh EH9 3FD, United Kingdom}
\affil[$2$]{School of Mathematical \& Computer Sciences and Maxwell Institute for Mathematical Sciences, Heriot-Watt University, Edinburgh EH14 4AP, United Kingdom}

\maketitle

\begin{abstract}
Rate‑induced tipping (``R-tipping") in ecological modelling is characterised by too-rapid change of an environmental parameter causing collapse of a population without any bifurcation being crossed (``B-tipping"). A well‑known example of a slow–fast predator-prey model in which R-tipping is observed [Vanselow, Wieczorek \& Feudel, Journal of Theoretical Biology, 479, 64--72 (2019)] suffers from unecological ``resurgence", whereby populations driven to functional extinction recover towards stable coexistence. We propose an analytically tractable model, incorporating a strong Allee effect into the prey dynamics in a slow-fast Lotka-Volterra-type predator-prey system, which induces bistability and thus renders the extinction state a genuine attractor.
Applying geometric singular perturbation theory (GSPT), we describe the dynamics of the extended model that is obtained by ``ramping" of the inverse prey carrying capacity. We show the presence of a parabolic-shaped, folded critical manifold which admits a folded saddle singularity, the strong canard of which separates solutions that track the moving coexistence state from those that tip to extinction. The analytical simplicity of our model allows us to derive explicit expressions for the critical rate that separates ``tracking” from ``tipping” dynamics. Finally, we argue that our system represents a minimal model for rate‑induced tipping to extinction, in that it incorporates three essential ingredients: bistability, a folded critical manifold, and a folded-saddle-type canard.
\end{abstract}

\section{Introduction}

Ecological systems can undergo sudden, and often irreversible, transitions from one
state to another in response to changes in external conditions. Such ``tipping points"
or critical transitions~\cite{Catastrophicshits}  are of central importance in ecology, where they are associated with the collapse of populations and a loss of biodiversity. Classically, tipping has been understood as a bifurcation-induced phenomenon (``B-tipping"), whereby a slowly varying parameter crosses a critical threshold at which an attracting state loses stability or disappears. More recently, however, it has been recognised that a system may fail to track a persistent stable state if an external parameter is varied too rapidly, which is the mechanism underlying rate-induced tipping (``R-tipping") \cite{ashwin_2017_parameter,ashwin_2012_tipping,  wieczorek_2010_excitability}. In contrast to B-tipping, R-tipping is triggered by the rate of change of a parameter exceeding a critical value, rather than by the parameter itself attaining a particular value; correspondingly, it cannot be detected by linear stability analysis or classical bifurcation theory.

An ecological example of R-tipping was given by Vanselow, Wieczorek,
and Feudel~\cite{vanselow_2019_when}, who studied the slow-fast Rosenzweig-MacArthur predator-prey
model \cite{rosenzweig1963graphical} with Holling type-II functional response and declining
habitat quality. In dimensionless form, their system is given by 
\begin{equation}
\begin{aligned}
    \varepsilon\frac{d u}{d\tau}&=u(1-\phi(\tau) u)-\frac{uv}{1+\eta u},\\
    \frac{d v}{d\tau}&=\frac{uv}{1+\eta u}-v,\\
    \frac{d\phi}{d\tau}&=r.
    \label{Vanselow}
\end{aligned}
\end{equation}
Here, the parameter $\phi$ is proportional to the inverse of the prey carrying capacity, with an increase in $\phi$ at some constant rate
$r$ modelling the dwindling of resources within the habitat due to a changing climate or habitat loss. In particular, it is shown in \cite{vanselow_2019_when}
that a slow decline in resource level can trigger the
collapse of a predator-prey population which is initiated at stable coexistence. For a fixed carrying capacity, with $\phi$ constant or, equivalently, $r=0$, the coexistence state in \eqref{Vanselow} is an attracting equilibrium when $\phi\in(\phi_{\rm min}, \phi_{\rm max})$, where $\phi_{\rm min}=\tfrac{\eta(1-\eta)}{1+\eta}$ and $\phi_{\rm max}=1-\eta$; hence, no bifurcation is encountered as $\phi$ is varied across that range. When $\frac{d\phi}{d\tau}=r$, i.e., as the carrying capacity is permitted to vary slowly, $\phi$ becomes a second slow variable in \eqref{Vanselow} alongside the
predator density $v$. The resulting reduced flow on the two-dimensional critical manifold for \eqref{Vanselow} admits a ``folded saddle" singularity at the fold of that critical manifold \cite{SZMOLYAN2001419}. The stable manifold of the folded
saddle is a separatrix that divides the attracting sheet of the critical manifold into two regions: initial states on one side of the separatrix track the moving coexistence state as $\phi$ increases, while those on the other side are drawn to the fold and ejected in
the fast prey direction, thus undergoing collapse. Whether a given trajectory ``tracks" or ``tips" is
hence determined not by the magnitude of $\phi$, but, rather, by its rate of change $r$.

For $r=0$ in Equation~\eqref{Vanselow}, the extinction state at $(u,v)=(0,0)$ is always unstable. Consequently, trajectories that tip do not converge to extinction; instead, after the prey density is driven to unrealistically low values, they are repelled once the invariant component $\{u=0\}$ of the critical manifold loses stability, resulting in resurgence, with populations of prey and predator ultimately returning to the moving coexistence state. The collapse is therefore only temporary, and genuine
extinction is recovered in \cite{vanselow_2019_when} by addition of noise. This resurgence is an artifact of the model structure rather than a reflection of the ecology of a collapsing ecosystem, in which vanishingly small populations are typically doomed to extinction.

\medskip

In this article, we present a simple modification of Equation~\eqref{Vanselow} which avoids resurgence, in the hope that it may be useful as a pedagogical, ``minimal" model of this type of R-tipping; specifically, we incorporate a strong Allee effect into the prey dynamics in \eqref{Vanselow}. The Allee effect \cite{allee} models the reduced per-capita growth of populations at low densities and introduces a critical threshold value ($A$) in \eqref{Vanselow} below which the prey population declines to zero. Crucially, the strong Allee effect renders the extinction state at the origin a genuine attractor, resulting in bistability. Unlike in \cite{vanselow_2019_when}, trajectories that tip are therefore driven permanently to extinction, with $(u,v)\to(0,0)$ as $\tau\to\infty$.

For analytical simplicity, and in line with our aim of minimality, we replace the Holling type-II functional response of \cite{vanselow_2019_when} with a Lotka–Volterra (Holling type-I) response, which is a reasonable assumption for moderate prey densities. As will become evident below, this simplification retains a polynomial prey nullcline while preserving the fast–slow folded geometry, the folded saddle singularity, and the bistability which, in combination, induce rate-induced tipping to extinction. Crucially, the resulting system is analytically tractable, which allows us to describe its reduced flow, locate the folded saddle singularity, approximate its stable manifold, and thereby derive an explicit leading-order expression for the critical rate $r_{\rm crit}$ that separates ``tracking" from ``tipping" dynamics in closed form, which we then confirm numerically.

\medskip

The remainder of the article is organised as follows. 
In Section~\ref{section:set-up}, we introduce a Lotka-Volterra model with strong Allee effect and type-I functional response, and we non-dimensionalise the model to recast it in slow-fast form. In Section~\ref{section:GSPT}, we apply geometric singular perturbation theory (GSPT) to show that, in a certain parameter regime characterised by the inverse carrying capacity, the extinction and coexistence states are both local attractors, resulting in bistability. In Section~\ref{section:ramped system}, we introduce the corresponding ``ramped" system, allowing the carrying capacity to decline slowly in time, and we show that rate-induced tipping can occur in the form of a folded saddle-type canard \cite{SZMOLYAN2001419, Canards}. In Section~\ref{section:4}, we prove that there exists a critical rate $r_{\rm crit}$ such that for $r<r_{\rm crit}$, the coexistence of predator and prey persists, whereas for $r>r_{\rm crit}$, rate-induced tipping results in the extinction of both populations. We conclude with a discussion in Section~\ref{section:discussion}. For the sake of readability, all mathematical proofs are relegated to an Appendix.

\section{Minimal model}
\label{section:set-up}

We consider a Lotka–Volterra model with strong Allee effect \cite[Equation~(1)]{VANVOORN2007451} as our minimal model,
\begin{equation}
    \begin{aligned}
    \frac{dU}{dT}&=RU\left(1-\tfrac{U}{K}\right)(U-A)-pUV,\\
    \frac{dV}{dT}&=cpUV-mV,
    \label{eq-dimensional}
    \end{aligned}
\end{equation}
where $U$ and $V$ are the respective prey and predator densities. The corresponding parameters are specified in Table~\ref{tab:1}. 
\begin{table}[ht]
\centering
\begin{tabular}{c l}
Symbol & Meaning \\
\hline
$R$ & Prey growth coefficient \\
$K$ & Prey carrying capacity  \\
$A$ & Allee threshold  \\
$p$ & Predation rate  \\
$c$ & Conversion efficiency \\
$m$ & Predator mortality rate  \\
\hline
\end{tabular}
\caption{Model parameters in Equation~\eqref{eq-dimensional}.}
\label{tab:1}
\end{table}
Throughout, we assume that $A<\tfrac{m}{cp}<K$, which implies that predator regulates prey at a density that is sufficiently high to preclude extinction due to the Allee effect, yet below the habitat's carrying capacity. Furthermore, we assume $0<c<1$ for the conversion efficiency of prey to predator, which reflects the fact that only a proportion of consumed prey biomass is converted into predator biomass \cite{rosenzweig1963graphical}.

We now perform a well-motivated change of coordinates which transforms \eqref{eq-dimensional} into a dimensionless formulation with an explicit slow-fast structure. Specifically, let 
\[
u=\tfrac{cp}{m}U,\quad v=\tfrac{cp^2}{mR}V,\quad \tau=mT,\quad \phi=\tfrac{m}{cpK},\quad a=\tfrac{cp}{m}A,\quad\text{and}\quad \varepsilon=\tfrac{cp}{R};
\]
then, a simple calculation shows that the dimensionless system of equations resulting from \eqref{eq-dimensional} reads
\begin{subequations}\label{eq-1}
\begin{align}
\varepsilon\dot{u} &=u(1-\phi u)(u-a)-uv, \\
\dot{v} &=v(u-1),
\end{align}
\end{subequations}
where $u$ and $v$ denote the dimensionless prey and predator populations, respectively, while the overdot indicates differentiation with respect to (the slow) time $\tau$. Moreover, $\phi$ is proportional to the inverse of the carrying capacity of prey, with $\phi\in(0,1)$ and $a\in(0,1)$ due to our earlier assumption that $A<\tfrac{m}{cp}<K$. Finally, we assume that the prey growth rate $R$ is large relative to the rate $cp$ of biomass gained by predator from prey, i.e., that $R\gg cp$, which implies that $\varepsilon=\tfrac{cp}{R}\ll 1$ is a small singular perturbation parameter.

Clearly, the positive quadrant $\mathbb{R}_+^2$ of $\mathbb{R}^2$ is invariant under the flow of \eqref{eq-1}.
In the next section, we apply GSPT \cite{FENICHEL197953} to show that Equation~\eqref{eq-1} exhibits bistability for $\varepsilon$ sufficiently small.

\section{Geometric analysis}
\label{section:GSPT}
Equation~\eqref{eq-1} is a singular perturbation problem in the standard slow formulation of GSPT; for $\varepsilon=0$, we obtain the reduced problem
\begin{subequations}\label{eq-2}
\begin{align}
0 &=u(1-\phi u)(u-a)-uv, \label{eq-2a} \\
\dot{v} &=v(u-1). \label{eq-2b}
\end{align}
\end{subequations}
Equation~\eqref{eq-2a} defines the so-called critical manifold $S_0$ which, in our case, consists of two sub-manifolds $S_0^0$ and $S_0^1$ that are defined by
\begin{subequations}\label{eq-6}
\begin{align}
S_0^0 &=\big\{(u,v)\in\mathbb{R}_+^2\,\big|\, u=0\big\}\quad\text{and} \label{eq-6a} \\
S_0^1 &=\big\{(u,v)\in\mathbb{R}_+^2\,\big|\, v=(1-\phi u)(u-a)\big\}, \label{eq-6b}
\end{align}
\end{subequations}
respectively. Introducing the fast time $t=\tau/\varepsilon$ in \eqref{eq-1}, we find the equivalent fast system
\begin{subequations}\label{eq-3}
\begin{align}
u' &=u(1-\phi u)(u-a)-uv, \\
v' &=\varepsilon v(u-1),
\end{align}
\end{subequations}
with the prime denoting differentiation with respect to $t$. The singular limit of $\varepsilon=0$ in Equation~\eqref{eq-3} yields the layer problem
\begin{subequations}\label{eq-4}
\begin{align}
u' &=u(1-\phi u)(u-a)-uv, \\
v' &=0
\end{align}
\end{subequations}
in which $v$ reduces to a parameter.

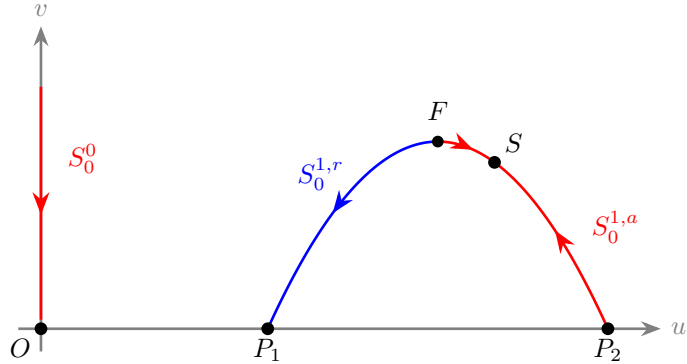
\begin{figure}[ht]
  \centering
\begin{tikzpicture}[>=Stealth,line width=1pt]


  \draw[->,gray] (-0.3,0) -- (8.2,0) node[right] {$u$};
  \draw[->,gray] (0,-0.3) -- (0,4.0) node[above] {$v$};

  \draw[red,flow=0.55] (0,3.2) -- (0,0.12);

  \draw[blue,flow=0.5]
     plot[domain=0.875:0.5,samples=60] ({\x*6},{(1-0.8*\x)*(\x-0.5)*22});

  \draw[red,flow=0.6]
     plot[domain=0.875:1.0,samples=40] ({\x*6},{(1-0.8*\x)*(\x-0.5)*22});
  \draw[red,flow=0.55]
     plot[domain=1.25:1.0,samples=40] ({\x*6},{(1-0.8*\x)*(\x-0.5)*22});

  \fill (5.25,2.475) circle (2.2pt);
  \node[above] at (5.25,2.62) {$F$};

  \fill (0,0)   circle (2.4pt) node[below left]  {$O$};
  \fill (3,0)   circle (2.4pt) node[below]       {$P_1$};
  \fill (6,2.2) circle (2.4pt) node[above right] {$S$};
  \fill (7.5,0) circle (2.4pt) node[below]       {$P_2$};

  \node[red]  at (0.55,2.25) {$S_0^{0}$};
  \node[blue] at (3.7,2.05)  {$S_0^{1,r}$};
  \node[red]  at (7.6,1.35)  {$S_0^{1,a}$};

\end{tikzpicture}
\caption{The singular geometry of \eqref{eq-1} for $a=\tfrac{1}{2}$ and $\phi=\tfrac{4}{5}$, which implies $\phi_{\rm min}=\tfrac{2}{3}$ and $\phi_{\rm max}=1$.}
\label{fig: singular geometry planar}
\end{figure}

The geometry of the critical manifold $S_0$ is illustrated in Figure~\ref{fig: singular geometry planar}. We have the following result on the normal hyperbolicity thereof.
\begin{lemma}\label{lem-1}
The critical manifold $S_0$ has the following properties.
\begin{enumerate}
\item The sub-manifold $S_0^0$ is normally attracting under the flow of Equation~\eqref{eq-4}.
\item The sub-manifold $S_0^1$ is normally repelling (normally attracting) for $u<\frac{1+a\phi}{2\phi}$ ($u>\frac{1+a\phi}{2\phi}$), losing hyperbolicity at the fold point $F:\ \big(\frac{1+a\phi}{2\phi},\frac{(1-a\phi)^2}{4\phi}\big)$.
\end{enumerate}
\end{lemma}
Henceforth, we denote the attracting (repelling) portion of $S_0^1$ by $S_0^{1,a}$ ($S_0^{1,r}$).
\begin{remark}
While the sub-manifolds $S_0^0$ and $S_0^1$ lose hyperbolicity in their point of intersection $(0,-a)$, that point is ecologically irrelevant and, hence, excluded from our definition of $S_0$.
\end{remark}

Next, we consider the reduced flow on the critical manifold $S_0$.
\begin{prop}\label{prop-1}
The reduced flow of Equation~\eqref{eq-2} can be written as
\begin{subequations}\label{eq-5}
\begin{align}
\dot{v} &=-v \qquad\text{on }S_0^0\text{ and} \label{eq-5a} \\
\dot{u} &=(u-a)\frac{(1-\phi u)(u-1)}{1+a\phi-2\phi u} \qquad\text{on }S_0^1, \label{eq-5b}
\end{align}
\end{subequations}
respectively.
\end{prop}
The equilibria of the reduced flow are determined from \eqref{eq-5}: on $S_0^0$, we find the extinction state $O:\ (0,0)$, while on $S_0^1$, we have the prey-only states $P_1:\ (a,0)$ and $P_2:\ (\phi^{-1},0)$ and the coexistence state $S:\ (1,(1-\phi)(1-a))$. To ensure the ecological relevance of $S$, we will henceforth assume $0<\phi< 1$; correspondingly, we define $\phi_{\rm max}\equiv 1$.
The stability of those equilibria is as follows; see again Figure~\ref{fig: singular geometry planar} for an illustration.
\begin{prop}\label{prop-2}
Let $a\in(0,1)$ be fixed, and let $\phi_{\rm min}:=(2-a)^{-1}\in(\frac{1}{2},1)$.
\begin{enumerate}
\item The extinction state $O$ is attracting under the reduced flow on $S_0^0$.
\item The prey-only state $P_1$ is attracting under the reduced flow on $S_0^1$.
\item The prey-only state $P_2$ is repelling under the reduced flow on $S_0^1$.
\item The coexistence state $S$ is repelling under the reduced flow on $S_0^1$ for $\phi\in(0,\phi_{\rm min})$ and attracting for $\phi\in(\phi_{\rm min},\phi_{\rm max})$.
\end{enumerate}
\end{prop}
\begin{remark}
\begin{enumerate}
\item For $\phi=1=a$, the manifold $S_0^1$ degenerates to $v=-(u-1)^2$, with the equilibria $P_i$ ($i=1,2)$ and $S$ merging with the fold point $F$ at $u=1$; that case is excluded by assumption, however.
\item The equilibria $P_i$ ($i=1,2$) are always distinct from $F$, as either can only coincide with $F$ for $\phi=a^{-1}>1$, which is again inadmissible.
\item As in \cite{vanselow_2019_when}, the coexistence state $S$ merges with $F$ and undergoes a (singular) Hopf bifurcation at $\phi=\phi_{\rm min}$; for $\phi=\phi_{\rm max}$, the equilibria $S$ and $P_2$ exchange stability in a transcritical bifurcation.
\end{enumerate}
\end{remark}

In the following, we will restrict to $\phi\in(\phi_{\rm min},\phi_{\rm max})$, see item 4 above, which implies that Equation~\eqref{eq-1} exhibits bistability, with both $O$ and $S$ being local attractors. In fact, since the vector field in \eqref{eq-1} is $\varepsilon$-independent, it follows that $O$, $P_i$ ($i=1,2)$, and $S$ are equally equilibria of Equation~\eqref{eq-1} -- or, equivalently, of \eqref{eq-3} -- for $\varepsilon$ positive and sufficiently small, as can be seen by intersection of the corresponding nullclines. Hence, these four states will be equilibria of the perturbed slow flow on persistent, normally hyperbolic portions of $S_0$ \cite{FENICHEL197953}. 

Importantly, the bistability of \eqref{eq-1} partitions the phase plane into the basins of attraction of the coexistence state $S$ and the extinction state $O$. In Section~\ref{section:4} below, we show how a slow decrease in the carrying capacity can drive a trajectory from the basin of $S$ into that of $O$, resulting in the collapse of both populations.

\section{Extended (ramped) system}
\label{section:ramped system}
As in \cite{vanselow_2019_when}, we now extend Equation~\eqref{eq-1} by defining $\phi$ to be a dynamic variable that increases linearly with rate $r$ provided $\phi\in(\phi_{\rm min},\phi_{\rm max})$; in fact, we assume $\phi\in(\phi_{{\rm min},\delta},\phi_{{\rm max},\delta})$ with $\phi_{\rm min, \delta}=\phi_{\rm min}+\delta$ and $\phi_{\rm max, \delta}=\phi_{\rm max}-\delta$, where $\delta$ is a small, but fixed, constant, to avoid the coexistence state undergoing a ``classical" bifurcation at $\phi=\phi_{\rm min}$ or $\phi=\phi_{\rm max}$.
The resulting, extended (``ramped") system of equations then reads
\begin{subequations}\label{eq-ramped system}
\begin{align}
\varepsilon\dot{u} &=u(1-\phi u)(u-a)-uv, \\
\dot{v} &=v(u-1), \\
\dot{\phi} &=\begin{cases}
r & \text{for }\phi\in[\phi_{{\rm min},\delta},\phi_{{\rm max},\delta}), \\
0 & \text{otherwise}
\end{cases}
\end{align}
\end{subequations}
in its slow formulation. Correspondingly, for $\varepsilon=0$, we find the reduced problem
\begin{subequations}\label{eq-8}
\begin{align}
0 &=u(1-\phi u)(u-a)-uv, \\
\dot{v} &=v(u-1), \\
\dot{\phi} &=\begin{cases}
r & \text{for }\phi\in[\phi_{{\rm min},\delta},\phi_{{\rm max},\delta}), \\
0 & \text{otherwise}.
\end{cases}
\end{align}
\end{subequations}
In the fast time $t=\tau/\varepsilon$, we obtain the equivalent fast system
\begin{subequations}\label{eq-ramped fast}
\begin{align}
u' &=u(1-\phi u)(u-a)-uv, \\
v' &=\varepsilon v(u-1), \\
\phi' &=\begin{cases}
\varepsilon r & \text{for }\phi\in[\phi_{{\rm min},\delta},\phi_{{\rm max},\delta}), \\
0 & \text{otherwise},
\end{cases}
\end{align}
\end{subequations}
which reduces to the layer problem
\begin{subequations}\label{eq-ramped layer}
\begin{align}
u' &=u(1-\phi u)(u-a)-uv, \\
v' &=0, \\
\phi' &=0
\end{align}
\end{subequations}
in the singular limit of $\varepsilon=0$. 
We denote the critical manifold of \eqref{eq-8} as $S_{0, \phi}$, which is now a two-dimensional surface due to $\phi$ being an additional slow variable.
The stability properties of $S_{0, \phi}$ are consistent with those of the critical manifold $S_0$ for \eqref{eq-1}, in that the stability of any point $(u,v)\in S_0$, for given $\phi$, will be identical to that of the point $(u,v,\phi)\in S_{0, \phi}$. Furthermore, the equilibria $O$, $P_1$, $P_2$, and $S$ of \eqref{eq-1} and the fold point $F$ become curves on $S_{0, \phi}$ in \eqref{eq-ramped system}. We denote these curves by $O_{\phi}$, $P_{1, \phi}$, $P_{2, \phi}$, $S_{\phi}$, and $F_{\phi}$, respectively, to indicate their dependence on $\phi$.

We focus on the extended sub-manifold $S_{0,\phi}^1$ here, as it contains the extension $S_\phi:\ (1,(1-\phi)(1-a),\phi)$ of the coexistence state $S$.
In analogy to Proposition~\ref{prop-1}, we have the following result.
\begin{prop}\label{prop-3}
The reduced flow of Equation~\eqref{eq-8} on $S_{0,\phi}^1$ can be written as
\begin{subequations}\label{eq-9}
\begin{align}
\dot{u} &=(u-a)\frac{(1-\phi u)(u-1)+ru}{1+a\phi-2\phi u}, \label{eq-9a} \\
\dot{\phi} &=r.
\end{align}
\end{subequations}
\end{prop}
The desingularised system corresponding to \eqref{eq-9} is obtained by a rescaling of time via $\frac{d}{ds}=-(1+a\phi-2\phi u)\frac{d}{d\tau}$ which removes the singularity of the reduced flow at the fold $F_{\phi}$ where the denominator in \eqref{eq-9a} vanishes: 
\begin{subequations}\label{eq-10}
\begin{align}
\frac{du}{ds} &=-(u-a)[(1-\phi u)(u-1)+ru], \label{eq-10a} \\
\frac{d\phi}{ds} &=-r(1+a\phi-2\phi u). \label{eq-10b}
\end{align}
\end{subequations}
Here, we emphasise that this rescaling preserves the directionality of the reduced flow on the attracting portion $S_{0,\phi}^{1,a}$ of $S_{0,\phi}^1$, as $1+a\phi-2\phi u<0$ there.
Folded singularities for Equation~\eqref{eq-ramped system} are defined as equilibria of the desingularised Equation~\eqref{eq-10} \cite{SZMOLYAN2001419}. We have the following result.
\begin{prop}\label{prop-4}
Let $r,\delta>0$ be sufficiently small, and let $\phi\in(\phi_{{\rm min},\delta},\phi_{{\rm max},\delta})$. Then, Equation~\eqref{eq-ramped system} admits a unique, ecologically relevant folded singularity $FS_{\phi,r}$ which satisfies the following properties.
\begin{enumerate}
\item The singularity 
\begin{align}\label{eq-13}
FS_{\phi,r}:\ \bigg(1-\frac{2-a}{1-a}r+\mathcal{O}(r^2),\,\frac{1}{2-a}+\frac{2}{(1-a)(2-a)}r+\mathcal{O}(r^2)\bigg)
\end{align}
is a regular perturbation of $S_\phi$, with $FS_{\phi,0}:\ (1,\phi_{\rm min})$.
\item The singularity $FS_{\phi,r}$ is a folded saddle equilibrium, with eigenvalues 
\begin{align}\label{eq-11}
\lambda_-=-\frac{(1-a)^2}{2-a}+\mathcal{O}(r)\quad\text{and}\quad \lambda_+=(2-a)r+\mathcal{O}(r^2)
\end{align}
and corresponding eigenvectors 
\begin{align}\label{eq-12}
\mathbf{v}_-=\bigg(1, -\frac{2}{(1-a)^2}r+\mathcal{O}(r^2)\bigg)^T\quad\text{and}\quad \mathbf{v}_+=\bigg(-\frac{(2-a)^2}{(1-a)^2}r+\mathcal{O}(r^2), 1\bigg)^T, 
\end{align}
respectively.    
\end{enumerate}
\end{prop}
The stable manifold $W^{\rm s}(FS_{\phi, r})$ of the folded saddle $FS_{\phi, r}$ is a singular (in $\varepsilon$) canard \cite{SZMOLYAN2001419} $\Gamma_0$: a special trajectory which crosses from $S_{0, \phi}^{1, a}$ to $S_{0, \phi}^{1, r}$. In the next section, we identify that canard as the threshold separating trajectories that track the moving coexistence state $S_\phi$ from those that tip to extinction, and we show that whether a given trajectory tracks or tips is determined by the rate $r$.

\section{Rate-induced tipping}
\label{section:4}

Next, we approximate the critical rate $r_{\rm crit}$ which separates trajectories that track the moving coexistence state $S_\phi$ with increasing $\phi$ from those that undergo tipping. To that end, we follow the procedure in Appendix F of \cite{vanselow_2019_when}, approximating the stable manifold $W^{\rm s}(FS_{\phi,r})$ of the folded saddle $FS_{\phi,r}$ by the corresponding eigendirection. Recall the expression for the folded singularity $FS_{\phi,r}$ and its stable eigenvector $\mathbf{v}_-$ in Equations~\eqref{eq-13} and \eqref{eq-12}, respectively; furthermore, as in \cite{vanselow_2019_when}, assume that $I=(u,v,\phi)(0)=(1,(1-\phi_{\rm min, \delta})(1-a), \phi_{\rm min, \delta})$, that is, the populations of prey and predator are initiated at the coexistence state $S_{\phi}$ for $\phi=\phi_{\rm min, \delta}$.
\begin{prop}
    For sufficiently small $r,\delta>0$, there exists a critical rate
    \begin{align}\label{eq-rcrit}
    r_{\rm crit}=\frac{(1-a)(2-a)}{2}\delta+\mathcal{O}(\delta^2)
    \end{align}
    such that the trajectory $\gamma_I$ of the reduced flow in \eqref{eq-8} with initial condition $I$ behaves as follows.
    \begin{enumerate}
        \item For $r<r_{\rm crit}$, the trajectory $\gamma_I$ remains on $S^{1, a}_{0, \phi}$ for $\phi\in(\phi_{\rm min,\delta},\phi_{\rm max,\delta})$.
        \item For $r=r_{\rm crit}$, $\gamma_I$ lies on the singular canard $\Gamma_0$ and reaches the fold line $F_{\phi}$ at $FS_{\phi,r}$ before crossing onto $S_{0, \phi}^{1, r}$.
        \item For $r>r_{\rm crit}$, $\gamma_I$ reaches the fold curve $F_{\phi}$ in finite time, where it leaves $S_{0, \phi}^1$ along a fast fibre under the layer flow of \eqref{eq-ramped layer}.
    \end{enumerate} 
    \label{prop:critical rate}
\end{prop}
\begin{remark}
    The prey component in the coexistence state $S_{\phi}$ is fixed at $u=1$ for all $\phi$, whereas the predator component $v=(1-\phi)(1-a)$ is strictly decreasing in $\phi$. Hence, the total biomass $u+v$ is maximised at $\phi=\phi_{\rm min}$, where $S_{\phi}$ intersects the fold line $F_{\phi}$. We therefore initiate the trajectory $\gamma_I$ at $\phi=\phi_{\rm min,\delta}$ to ensure that $S_{ \phi}$ lies on the normally attracting part of $S_{0, \phi}^{1}$, while remaining as close as possible to the point of maximal population biomass.
\end{remark}

We emphasise that, in Proposition~\ref{prop:critical rate}, we consider the reduced flow, with $\varepsilon=0$ in \eqref{eq-ramped system}, for which the singular canard $\Gamma_0$ forms the threshold between ``tracking" (item 1) and ``tipping" (item 3). When $\varepsilon>0$, the singular canard $\Gamma_0$ perturbs to a maximal canard $\Gamma_\varepsilon$ \cite{SZMOLYAN2001419}.

\begin{figure}[H]
    \centering
    \includegraphics[scale = 0.7]{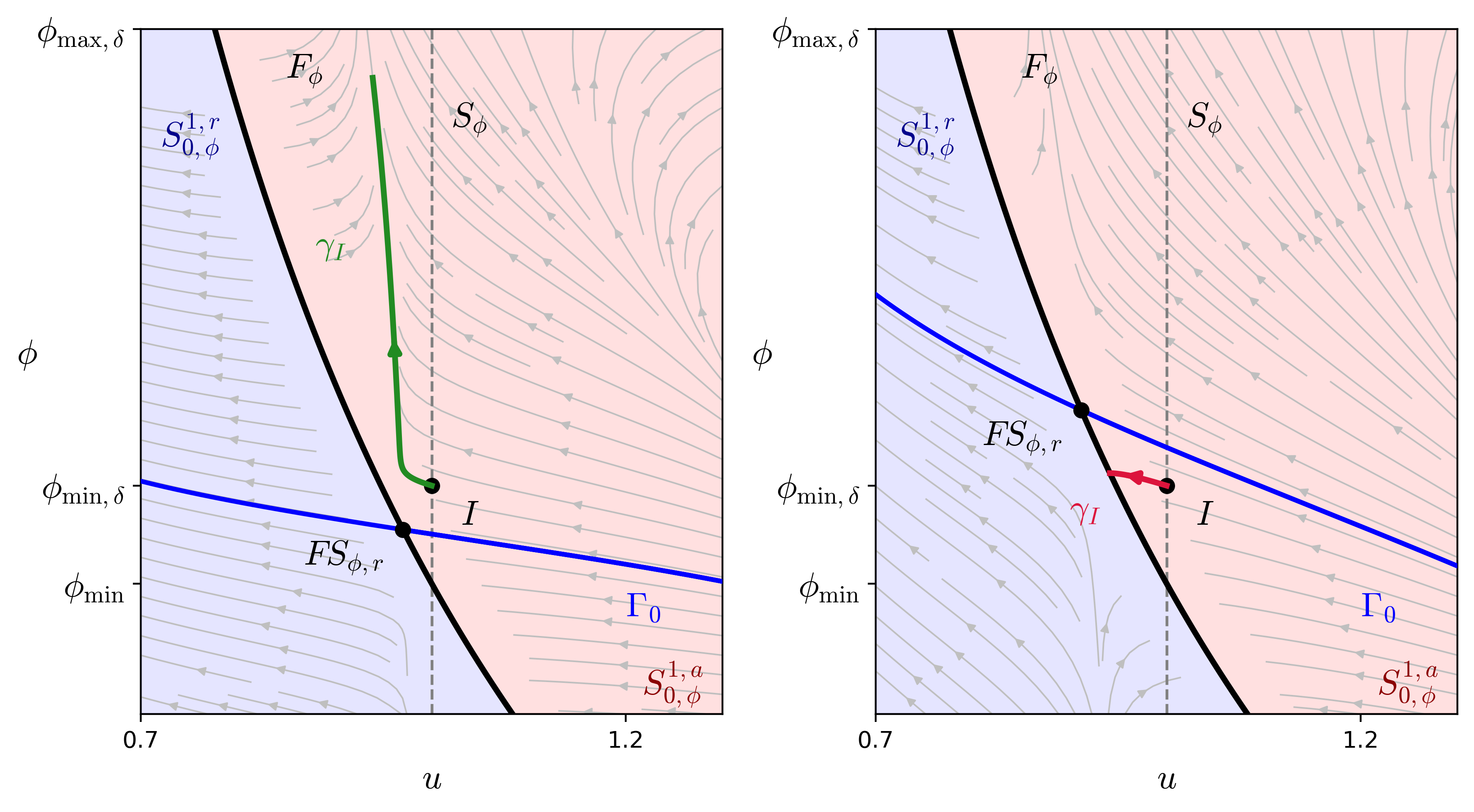}
    \caption{The dynamics of the projected reduced flow in \eqref{eq-9} on $S_{0,\phi}^1$ for $a=0.5$ and $\delta=0.05$, which implies $\phi_{\min}=(2-a)^{-1}=\tfrac{2}{3}\approx0.667$ and
    $r_{\rm crit}\approx\tfrac{(1-a)(2-a)}{2}\delta=0.01875$, by \eqref{eq-rcrit}. Left panel: $r=0.01$; right panel: $r=0.03$. The attracting sheet $S_{0,\phi}^{1,a}$ (in red) and the repelling sheet $S_{0,\phi}^{1,r}$ (in blue) meet along the fold curve
    $F_{\phi}$ (in black), where $u_F=\tfrac{1+a\phi}{2\phi}$.
    The singular canard $\Gamma_0$ is plotted in dark blue; the initial condition
    $I=(1,(1-\phi_{\min,\delta})(1-a),\phi_{\min,\delta})$, with $\phi_{\min,\delta}=\phi_{\min}+\delta\approx0.7167$, lies on
    the extended coexistence state $S_{\phi}: (1,(1-\phi)(1-a),\phi)$.}
    \label{fig-desingularised}
\end{figure}

In Figure~\ref{fig-desingularised}, we illustrate the result of Proposition~\ref{prop:critical rate} by plotting the dynamics of the projected reduced flow of \eqref{eq-ramped system} on $S_{0,\phi}^1$ in the two cases where $r<r_{\rm crit}$ and $r>r_{\rm crit}$. Specifically for $r>r_{\rm crit}$, the trajectory $\gamma_I$ (in red) reaches the fold line $F_{\phi}$ before leaving along a fast fibre under the layer flow of \eqref{eq-ramped layer}.

\begin{figure}[H]
    \centering
    \includegraphics[scale=0.2]{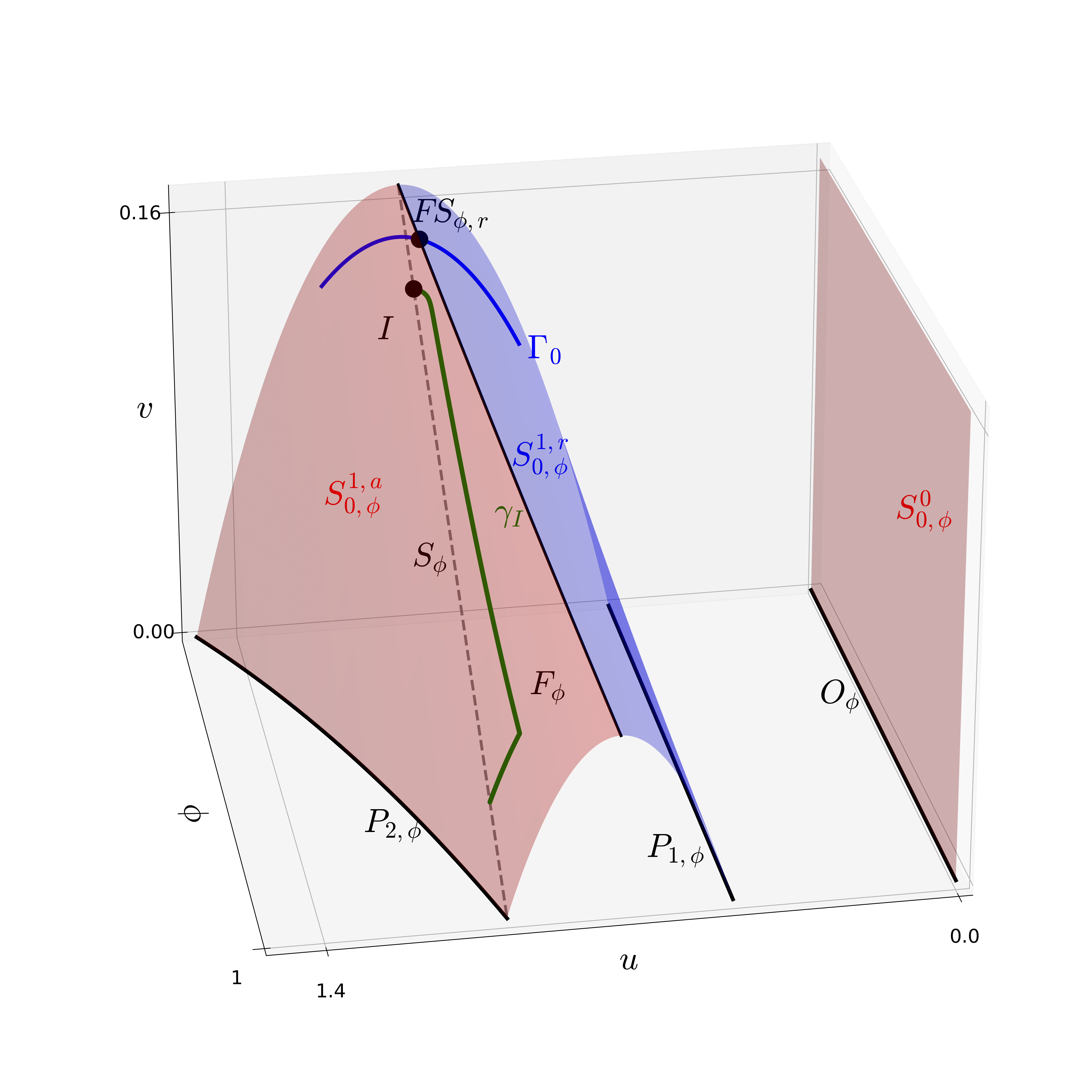}
    \includegraphics[scale=0.2]{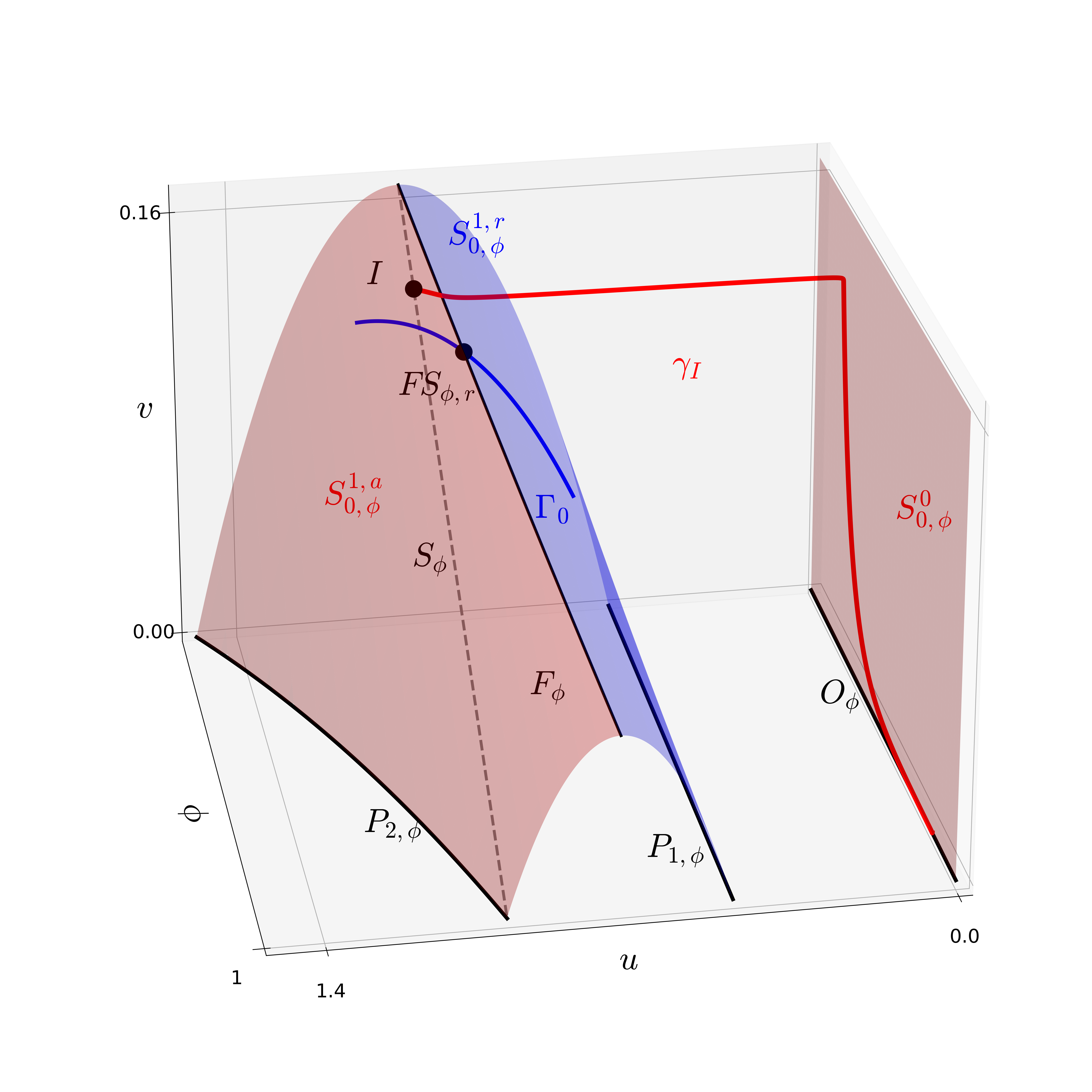}
    \caption{The dynamics of \eqref{eq-ramped system} with $a=0.5$, $\delta=0.05$, and $\varepsilon=0.001$, which implies $r_{\rm crit}\approx \tfrac{(1-a)(2-a)}{2}\delta=0.01875$, by \eqref{eq-rcrit}. Left panel: tracking ($r=0.01$); right panel: tipping ($r=0.03$). The singular canard $\Gamma_0$ (in dark blue) is approximated via the stable eigendirection of $FS_{\phi, r}$; see Proposition~\ref{prop-4}.}
    \label{fig-fullsystem}
\end{figure}

By the bistability established in Section~\ref{section:GSPT}, a trajectory that leaves $S_{0, \phi}^{1, a}$ along a fast fibre of \eqref{eq-ramped layer} enters the basin of attraction of $S_{0, \phi}^0$; the reduced flow thereon then gives convergence to $O_{\phi}$ due to $\dot{v}=-v$. Hence, for $r>r_{\rm crit}$, rate-induced tipping drives both populations permanently to extinction: $(u,v)\to(0,0)$ as $\tau\to\infty$, which is in contrast to \cite{vanselow_2019_when}, where the collapse is only temporary.
Figure~\ref{fig-fullsystem} confirms this prediction for the full ramped system, Equation~\eqref{eq-ramped system}: for $r=0.01<r_{\rm crit}\approx0.01875$, by the leading-order approximation in Equation~\eqref{eq-rcrit}, the trajectory $\gamma_I$ tracks the coexistence state $S_\phi$, whereas for $r=0.03>r_{\rm crit}$, tipping occurs, with $\gamma_I\to O_\phi$ and both $u$ and $v$ becoming extinct.

\begin{figure}[H]
    \centering
    \includegraphics[scale=0.6]{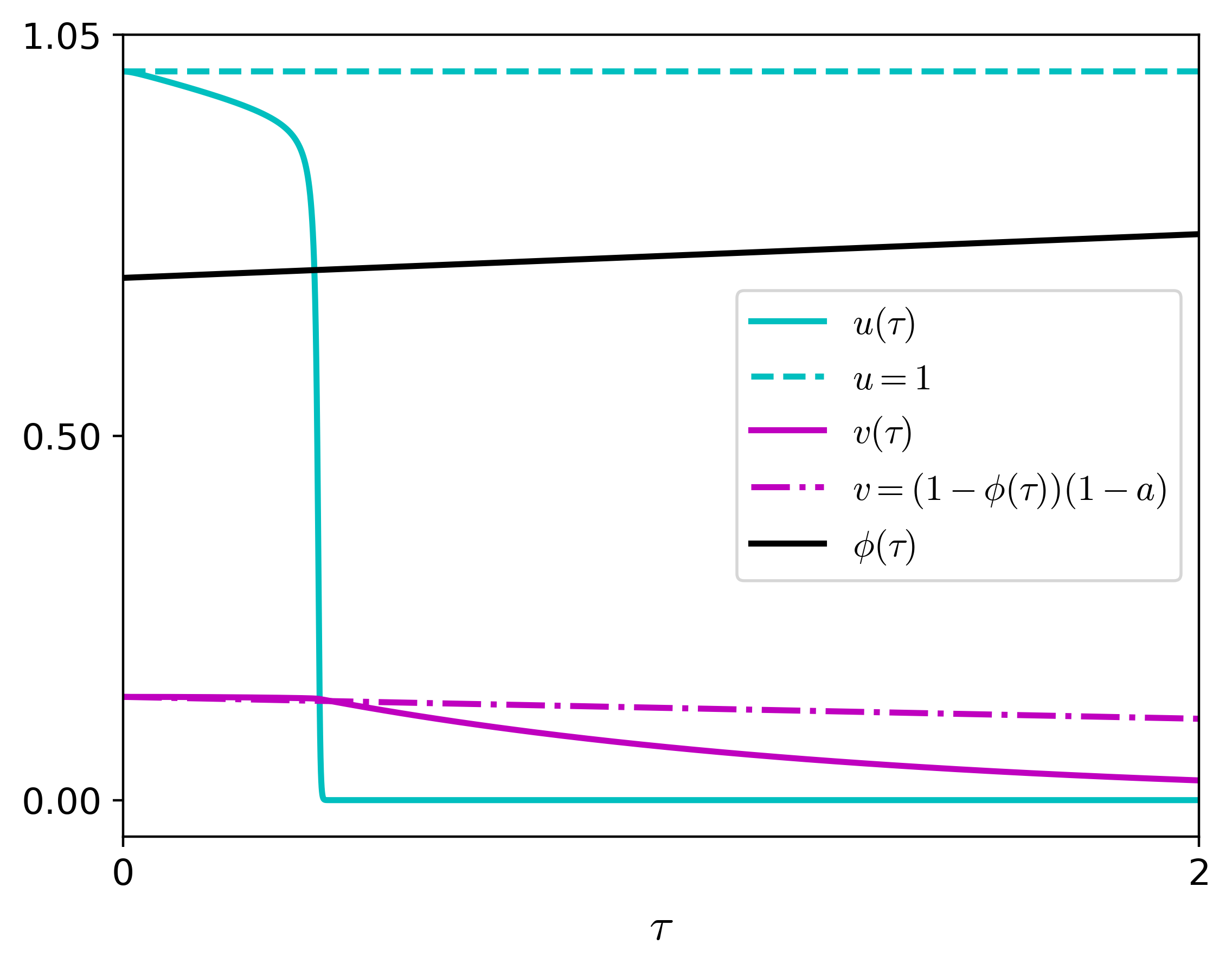}
    \caption{Time series of the trajectory $\gamma_I$ illustrated in Figure~\ref{fig-fullsystem} (left panel; $r=0.03$) in the slow time $\tau$. Here, $u$, $v$, and $\phi$ are represented in cyan, magenta, and black, respectively. A comparison against the evolution of the stable coexistence state $S_\phi$ is also given (dashed; dash-dotted).}
    \label{fig:timeseries}
\end{figure}

In Figure~\ref{fig:timeseries}, we compare the time series of the tipping trajectory illustrated in Figure~\ref{fig-fullsystem} to that of the extended coexistence state $S_{\phi}=(1, (1-\phi(\tau))(1-a), \phi(\tau))$. Standard linear theory predicts that any trajectory which is initiated sufficiently closely to $S_{\phi}$ will be attracted towards it for any fixed $\phi\in[\phi_{\rm min,\delta}, \phi_{\rm max,\delta})$. However, the underlying canard phenomenon allows for such trajectories to ``tip" towards $O_{\phi}$ instead.

To verify our approximation for $r_{\rm crit}$ numerically, we refine the result of Proposition~\ref{prop:critical rate} by providing the next-order term in $\delta$ in the expansion in Equation~\eqref{eq-rcrit}:
\begin{cor}
For sufficiently small $r,\delta>0$, the critical rate $r_{\rm crit}$ defined in Proposition~\ref{prop:critical rate} admits the expansion
    \begin{equation}
        r_{\rm crit}=\frac{(1-a)(2-a)}{2}\delta+\frac{(1-a)(2-a)^3}{2}\delta^2+\mathcal{O}(\delta^3).
        \label{eq-rcrit-order2}
    \end{equation}
    \label{cor:critical rate}
\end{cor}

\begin{figure}
    \centering
    \includegraphics[scale=0.575]{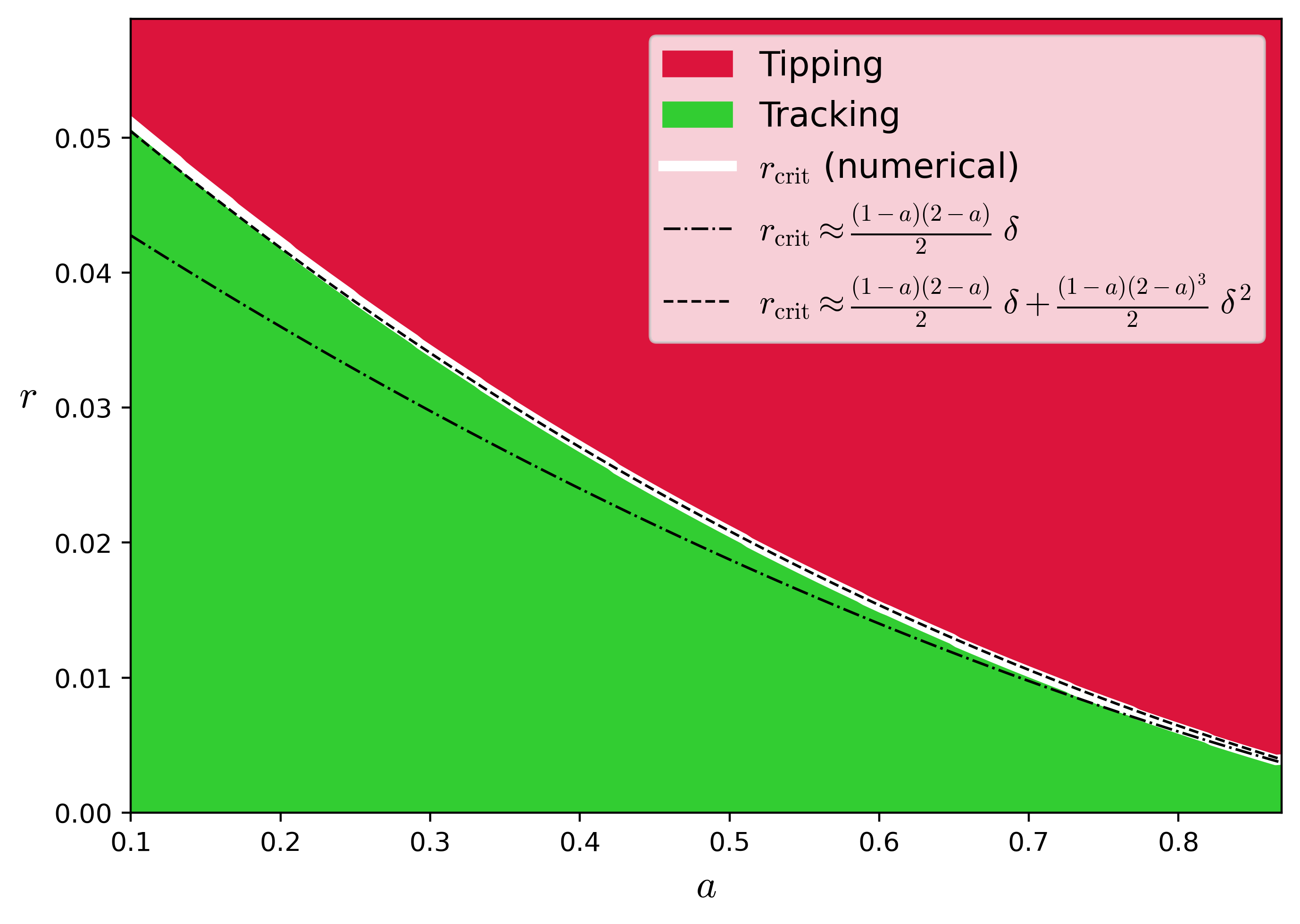}
    \caption{The numerical value of the critical rate $r_{\rm crit}$ (in white) as a function of $a$ for $\delta=0.05$ and $\varepsilon=0.001$ in comparison to the leading-order (dash-dotted black) and second-order (dashed black) expansion of $r_{\rm crit}$ in $\delta$; cf.~\eqref{eq-rcrit} and \eqref{eq-rcrit-order2}, respectively.}
    \label{fig:rcrit}
\end{figure}

In Figure~\ref{fig:rcrit}, we illustrate the numerical value of $r_{\rm crit}$, which is computed via a bisection method that is applied to the function
\begin{equation}
    \beta(r)=\begin{cases}
+1 & \text{if }\gamma_I\text{ remains in a neighborhood of } S_{0,\phi}^{1, a} \text{ for }  \phi<\phi_{\rm max, \delta}; \\
-1 & \text{if }\gamma_I\text{ leaves a neighborhood of } S_{0,\phi}^{1, a} \text{ for }  \phi<\phi_{\rm max, \delta}; \\
0 & \text{if }\gamma_I \text{ lies on the strong canard } \Gamma_{\varepsilon} \text{ for }  \phi<\phi_{\rm max, \delta};
\end{cases}
\end{equation}
here, $\delta=0.05$ and $\varepsilon=0.001$.
We set the tolerance of the bisection method to $10^{-5}$. Figure~\ref{fig:rcrit} indicates that the second-order expansion for $r_{\rm crit}$ in \eqref{eq-rcrit-order2} is in good agreement with the numerically computed value throughout, while the leading-order expansion in \eqref{eq-rcrit} performs well for $a$ close to $1$. That is not too surprising, as the coefficients of $\delta$ and $\delta^2$ are zero for $a=1$; cf.~Equation~\eqref{eq-rcrit-order2}. Correspondingly, we see that tipping dynamics becomes more prevalent with increasing $a$, which can also be understood from Equation~\eqref{eq-13}: as $a\to 1^-$, we require $r\to 0^+$ for the coordinates of $FS_{\phi,r}$ to remain bounded, which implies that trajectories will by default tip in accordance with item 3 in Proposition~\ref{prop:critical rate}.

\section{Discussion}
\label{section:discussion}

We have presented a slow-fast Lotka-Volterra-type predator-prey model in which a strong Allee effect in the prey equation converts the temporary, rate-induced collapse observed in \cite{vanselow_2019_when} into permanent extinction, thus preventing resurgence. In the reduced system, Equation~\eqref{eq-2}, i.e., for $\varepsilon=0$, the coexistence state $S$ and the extinction state $O$ are simultaneously attracting
(Proposition~\ref{prop-2}), which implies that \eqref{eq-1} exhibits bistability. Ramping of the inverse carrying capacity $\phi$ at a rate $r$ converts $\phi$ into an additional dynamic (slow) variable and generates
a folded saddle singularity $FS_{\phi,r}$ (Proposition~\ref{prop-4}) whose stable manifold is a singular canard. This canard constitutes the threshold separating trajectories that ``track" the moving coexistence state $S_\phi$ from those that tip towards the extinction state $O_\phi$. Correspondingly, as the threshold is crossed, trajectories leave the basin of attraction of $S_{\phi}$ and enter that of $O_{\phi}$. Due to the analytical simplicity of our Lotka-Volterra-type model, we were able to locate the folded saddle singularity, approximate its stable manifold, and thereby obtain the explicit leading-order critical rate $r_{\rm crit}$ (Proposition~\ref{prop:critical rate}) explicitly. Finally, we confirmed our findings by numerical simulation of the ``full" model, Equation~\eqref{eq-1}; cf.~Figure~\ref{fig-fullsystem}.

Finally, we argue that \eqref{eq-1} represents a minimal predator-prey system that exhibits rate-induced tipping to permanent extinction, with the populations of both prey and predator converging to a true extinction state without recovering. The three essential, relevant features of the model are: bistability, with both the extinction state and the coexistence state being local attractors throughout; a folded, parabolic-shaped critical manifold which loses normal hyperbolicity along a simple fold curve; and a folded saddle singularity whose strong canard corresponds to the tipping threshold. (Here, we remark that an analytically more complex, three-dimensional model of a reef ecosystem similarly exhibits resurgence-free tipping which is mediated through a folded node singularity; see \cite{antidze2026rateinducedtippingcoralreef} for details.)

In particular, we note that our simplification of the functional response from type II to type I in \eqref{Vanselow} -- i.e., the removal of saturation in the predation rate at high prey density -- would \textit{a priori} result in normal hyperbolicity of the entire non-trivial sub-manifold of the critical manifold, which would therefore admit no folded singularities. However, the simultaneous inclusion of a strong Allee effect in the prey equation in \eqref{eq-1} preserves the parabolic shape and folded geometry of our critical manifold and, hence, guarantees the presence of both a fold curve and a folded singularity, which is always of folded saddle type in the regime considered here. The change in geometry of the prey nullcline which is due to the Allee effect also ensures bistability, since the trivial (``extinction") sub-manifold of the critical manifold no longer intersects the non-trivial sub-manifold at positive predator densities, as is the case for \eqref{Vanselow}: the extinction state becomes a true attractor, which precludes resurgence due to the ``delayed loss of stability" on the trivial sub-manifold that is observed in \cite{vanselow_2019_when}.
The simplification that ensues in the predator equation in \eqref{eq-1}, on the other hand, allows us to locate the folded saddle singularity, and determine the critical rate that separates ``tracking" from ``tipping" behaviour, in closed form. We verify that rate numerically, and we discuss its dependence on the choice of Allee threshold in the prey equation.

Lastly, if a Holling type-II functional response is retained in Equation~\eqref{eq-1}, the non-trivial sub-manifold of the corresponding critical manifold is cubic, rather than quadratic, but still admits a unique, ecologically relevant fold point separating the attracting and repelling portions of that sub-manifold; furthermore, the trivial (extinction) sub-manifold remains normally attracting. One hence again observes bistability in suitably defined parameter regimes, with the unique coexistence state and the extinction state being attracting. The corresponding extended (ramped) system admits a unique folded saddle singularity for sufficiently small rates $r$, as before, the strong canard of which acts as a threshold for tipping to extinction. Hence, the resulting dynamics is qualitatively equivalent to that described here for \eqref{eq-1}; however, the requisite algebra is more involved.

In summary, we hope that our ``prototypical" model for resurgence-free population collapse in slow-fast predator-prey systems can serve as both a pedagogically useful example and as a basis for further investigation of canard-mediated rate-induced tipping phenomena in ecological modelling.

\section*{Acknowledgements}
The authors would like to thank Andy White (Heriot-Watt University) for numerous thought-provoking discussions. BH and ZS were supported by the EPSRC Centre for Doctoral Training in Mathematical Modelling, Analysis
and Computation (MAC-MIGS), funded by the UK Engineering and Physical Sciences Research
Council (grant EP/S023291/1, Heriot-Watt University and the University of Edinburgh).

\addcontentsline{toc}{section}{References}
\printbibliography

@article {Canards,
    AUTHOR = {Wechselberger, Martin},
     TITLE = {{\`A} propos de canards ({A}propos canards)},
   JOURNAL = {Transactions of the American Mathematical Society},
    VOLUME = {364},
      YEAR = {2012},
    NUMBER = {6},
     PAGES = {3289--3309},
      ISSN = {0002-9947,1088-6850},
   MRCLASS = {34E17 (34C40 34E13)},
  MRNUMBER = {2888246},
MRREVIEWER = {Kiyoyuki\ Tchizawa},
       DOI = {10.1090/S0002-9947-2012-05575-9},
       URL = {https://doi.org/10.1090/S0002-9947-2012-05575-9},
}

@book {MTSD,
    AUTHOR = {Kuehn, Christian},
     TITLE = {Multiple time scale dynamics},
    SERIES = {Applied Mathematical Sciences},
    VOLUME = {191},
 PUBLISHER = {Springer, Cham},
      YEAR = {2015},
     PAGES = {xiv+814},
      ISBN = {978-3-319-12315-8; 978-3-319-12316-5},
   MRCLASS = {34-02 (34-01 34Exx 37-02 37C10 37Gxx)},
}

@article{vanselow_2019_when,
  author = {Vanselow, Anna and Wieczorek, Sebastian and Feudel, Ulrike},
  pages = {64-72},
  title = {When very slow is too fast -- collapse of a predator-prey system},
  doi = {10.1016/j.jtbi.2019.07.008},
  url = {https://www.sciencedirect.com/science/article/pii/S0022519319302851},
  volume = {479},
  year = {2019},
  journal = {Journal of Theoretical Biology}
}

@article{wieczorek_2010_excitability,
  author = {Wieczorek, Sebastian and Ashwin, Peter and Luke, Catherine M. and Cox, Peter M.},
  pages = {1243-1269},
  title = {Excitability in ramped systems: the compost-bomb instability},
  doi = {10.1098/rspa.2010.0485},
  volume = {467},
  year = {2011},
  journal = {Proceedings of the Royal Society A: Mathematical, Physical and Engineering Sciences}
}

@article{ashwin_2012_tipping,
  author = {Ashwin, Peter and Wieczorek, Sebastian and Vitolo, Renato and Peter Timothy Cox},
  pages = {1166-1184},
  publisher = {Royal Society},
  title = {Tipping points in open systems: bifurcation, noise-induced and rate-dependent examples in the climate system},
  doi = {10.1098/rsta.2011.0306},
  volume = {370},
  year = {2012},
  journal = {Philosophical Transactions of the Royal Society A}
}

@article{ashwin_2017_parameter,
  author = {Ashwin, Peter and Perryman, Clare and Wieczorek, Sebastian},
  pages = {2185-2210},
  title = {Parameter shifts for nonautonomous systems in low dimension: bifurcation- and rate-induced tipping},
  doi = {10.1088/1361-6544/aa675b},
  volume = {30},
  number = {6},
  year = {2017},
  journal = {Nonlinearity}
}

@article{SZMOLYAN2001419,
title = {Canards in $\mathbb{R}^3$},
journal = {Journal of Differential Equations},
volume = {177},
number = {2},
pages = {419-453},
year = {2001},
issn = {0022-0396},
doi = {https://doi.org/10.1006/jdeq.2001.4001},
url = {https://www.sciencedirect.com/science/article/pii/S002203960194001X},
author = {Peter Szmolyan and Martin Wechselberger}
}

@article{Catastrophicshits,
author = {Scheffer, Marten and Carpenter, Stephen and Foley, Jonathan A. and Folke, Carl and Walker, Brian},
year = {2001},
pages = {591-596},
title = {Catastrophic shifts in ecosystems},
volume = {413},
number = {6856},
journal = {Nature}
}

@article{allee,
author = {Allee, Warder C. and Bowen, Edith S.},
title = {Studies in animal aggregations: Mass protection against colloidal silver among goldfishes},
journal = {Journal of Experimental Zoology},
volume = {61},
number = {2},
pages = {185-207},
doi = {https://doi.org/10.1002/jez.1400610202},
url = {https://onlinelibrary.wiley.com/doi/abs/10.1002/jez.1400610202},
eprint = {https://onlinelibrary.wiley.com/doi/pdf/10.1002/jez.1400610202},
year = {1932}
}

@article{VANVOORN2007451,
title = {Heteroclinic orbits indicate overexploitation in predator–prey systems with a strong Allee effect},
journal = {Mathematical Biosciences},
volume = {209},
number = {2},
pages = {451-469},
year = {2007},
issn = {0025-5564},
doi = {https://doi.org/10.1016/j.mbs.2007.02.006},
url = {https://www.sciencedirect.com/science/article/pii/S0025556407000363},
author = {George A.K. {van Voorn} and Lia Hemerik and Martin P. Boer and Bob W. Kooi}
}

@article{FENICHEL197953,
title = {Geometric singular perturbation theory for ordinary differential equations},
author = {Neil Fenichel},
journal = {Journal of Differential Equations},
volume = {31},
number = {1},
pages = {53-98},
year = {1979},
issn = {0022-0396},
doi = {https://doi.org/10.1016/0022-0396(79)90152-9},
url = {https://www.sciencedirect.com/science/article/pii/0022039679901529}
}

@misc{antidze2026rateinducedtippingcoralreef,
      title={Rate-induced tipping in a coral reef ecosystem: A slow increase in fishing effort can induce reef collapse}, 
      author={Irakli Antidze and Brian Hennessy and Nikola Popovi\'c and Zak Sattar},
      year={2026},
      eprint={2605.23487},
      archivePrefix={arXiv},
      primaryClass={math.DS},
      url={https://arxiv.org/abs/2605.23487}, 
}

@article{rosenzweig1963graphical,
  title={Graphical representation and stability conditions of predator-prey interactions},
  author={Rosenzweig, Michael L. and MacArthur, Robert H.},
  journal={The American Naturalist},
  volume={97},
  number={895},
  pages={209-223},
  year={1963},
  publisher={Science Press}
}

\appendix

\section{Mathematical proofs}
\label{section:appendix:Proofs}

Proofs of all mathematical statements in the main text can be found below.

\subsection{Proof of Lemma~\ref{lem-1}}

Normal hyperbolicity of $S_0$ is decided by linearisation of the layer problem, Equation~\eqref{eq-4}: since
\[
\frac{\partial}{\partial u}(u')=(1-\phi u)(u-a)+u(1+a\phi-2\phi u)-v,
\]
$\frac{\partial}{\partial u}(u')\big|_{S_0^0}=-a-v<0$, together with $v\geq 0$, implies that $S_0^0$ is normally attracting throughout.

Similarly, $\frac{\partial}{\partial u}(u')\big|_{S_0^1}=u(1+a\phi-2\phi u)$, in combination with $u\geq a>0$ on the ecologically relevant portion of $S_0^1$, implies that $S_0^1$ is normally repelling (normally attracting) for $u<\frac{1+a\phi}{2\phi}$ $\big(u>\frac{1+a\phi}{2\phi}\big)$, as claimed.

Finally, $S_0^1$ loses normal hyperbolicity at $u=\frac{1+a\phi}{2\phi}$, which corresponds to the global maximum of the parabola $v=(1-\phi u)(u-a)$, yielding the fold point $F$. 

\subsection{Proof of Proposition~\ref{prop-1}}

The reduced flow under Equation~\eqref{eq-4} on $S_0^0$ is obtained for $u=0$ in \eqref{eq-2b}, which immediately gives \eqref{eq-5a}.

As the parabolic sub-manifold $S_0^1$ does not allow a global graph representation in $v$, we express the reduced flow thereon in $u$ \cite[Chapter $3$]{MTSD}. To that end, we differentiate \eqref{eq-6b} with respect to the slow time $\tau$, which gives 
\[
\dot{v}=-\phi\dot{u}(u-a)+(1-\phi u)\dot{u}.
\]
Then, we substitute $\dot{v}=(1-\phi u)(u-a)(u-1)$, by \eqref{eq-2b} and \eqref{eq-6b}, and solve for $\dot{u}$, which yields \eqref{eq-5b}.

\subsection{Proof of Proposition~\ref{prop-2}}

Item 1 follows immediately from Equation~\eqref{eq-5a}.

To prove items 2 through 4, we consider the sign of $\dot{u}$ in \eqref{eq-5b}: first, we note that the denominator $1+a\phi-2\phi u$ is negative (positive) for $u>\frac{1+a\phi}{2\phi}$ $\big(u<\frac{1+a\phi}{2\phi}\big)$.

For $a<u<\min\big\{1,\frac{1+a\phi}{2\phi}\big\}$, we have $u-1<0$; also, $1-\phi u>0$ due to $\phi<a^{-1}$ always, which implies, in sum, that $\dot{u}<0$, showing item 2. (We note that $1<\frac{1+a\phi}{2\phi}$ $\big(1>\frac{1+a\phi}{2\phi}\big)$ for $\phi<\phi_{\rm min}$ ($\phi>\phi_{\rm min}$).)

Similarly, for $\phi^{-1}>u>\max\big\{1,\frac{1+a\phi}{2\phi}\big\}$, we have $u-a>0$, as well as again $1-\phi u>0$, which equally yields $\dot{u}<0$, implying item 3.

Finally, we prove item 4. For $\phi\in(0,\phi_{\rm min})$, we have $\dot{u}<0$ when $a<u<1$ due to $u-a>0$, $1-\phi u>0$, $u-1<0$, and $1+a\phi-2\phi u>0$; similarly, $\dot{u}>0$ when $1<u<\frac{1+a\phi}{2\phi}$, as $u-1>0$ now. For $\phi\in(\phi_{\rm min},\phi_{\rm max})$, on the other hand, we have $\dot{u}>0$ when $\frac{1+a\phi}{2\phi}<u<1$ due to $u-a>0$, $1-\phi u>0$, $u-1<0$, and $1+a\phi-2\phi u<0$, as well as $\dot{u}<0$ for $1<u<\phi^{-1}$, since $u-1>0$ then.

\subsection{Proof of Proposition~\ref{prop-3}}

The proof is analogous to that of Proposition~\ref{prop-1}, with the modification that $\phi$ is $\tau$-dependent now, which gives
\[
\dot{v}=-(\dot{\phi}u+\phi\dot{u})(u-a)+(1-\phi u)\dot{u}
\]
for the $\tau$-derivative of \eqref{eq-6b}.

\subsection{Proof of Proposition~\ref{prop-4}}

Folded singularities for \eqref{eq-ramped system} correspond to equilibria of the desingularised, reduced system in \eqref{eq-10}; by \eqref{eq-10b}, these satisfy $\phi^\ast=(2u^\ast-a)^{-1}$. Then, \eqref{eq-10a} implies either
$u^\ast-a=0$ or $(1-\phi^\ast u^\ast)(u^\ast-1)+ru^\ast=0$. Hence, either $u^\ast=a$ or
\[
u^\ast=\frac{1+a(1+r)\pm\sqrt{1-2a(1+3r)+a^2(1+r)^2}}{2(1+2r)}.
\]
In the former case, $\phi^\ast=a^{-1}>1-\delta=\phi_{{\rm max},\delta}$ due to $a\in(0,1)$ and appropriately chosen $\delta$, which makes that equilibrium irrelevant. 

In the latter case, we see that, for $r$ positive and sufficiently small, the negative square root yields
$u^\ast=a+\frac{a^2}{1-a}r+\mathcal{O}(r^2)(>a)$. Then, $\phi^\ast>\phi_{{\rm max},\delta}$, as before, since $2u^\ast-a=a+2\frac{a^2}{1-a}r+\mathcal{O}(r^2)<1$ for sufficiently small $r$; hence, that equilibrium is equally irrelevant.
The positive square roots gives
\[
u^\ast=1-\frac{2-a}{1-a}r+\mathcal{O}(r^2),
\]
which is admissible due to $2u^\ast-a=2-a-2\frac{2-a}{1-a}r+\mathcal{O}(r^2)>1$ for sufficiently small $r$. Furthermore, 
\[
\phi^\ast=\frac{1}{2-a}+\frac{2}{(1-a)(2-a)}r+\mathcal{O}(r^2)>\phi_{{\rm min},\delta}
\]
for appropriately chosen $\delta$, which proves item 1.

To show item 2, we calculate the Jacobian of Equation~\eqref{eq-10} about $FS_{\phi,r}$:
\[
J\big|_{FS_{\phi,r}}=\begin{pmatrix}
-(u^\ast-a)(1+\phi^\ast-2\phi^\ast u^\ast+r) & (u^\ast-a)u^\ast(u^\ast-1) \\
2r\phi^\ast & -r(a-2u^\ast)
\end{pmatrix}\Bigg|_{FS_{\phi,r}}.
\]
Then,
\[
\det J\big|_{FS_{\phi,r}}=r(u^\ast-a)\big[(1+\phi^\ast-2\phi^\ast u^\ast+r)(a-2u^\ast)-2\phi^\ast u^\ast(u^\ast-1)\big] \\
=-r\big[(1-a)^2+\mathcal{O}(r)\big]
\]
where we have used $1-2\phi^\ast u^\ast=-a\phi^\ast$ and $a-2u^\ast=-\frac{1}{\phi^\ast}$, by \eqref{eq-10b}, as well as the expansions (in $r$) for $u^\ast$ and $\phi^\ast$ from above. Since $\det J|_{FS_{\phi,r}}<0$ for $r$ sufficiently small, $FS_{\phi,r}$ is a folded saddle.

A straightforward calculation shows that the eigenvalues of $ J\big|_{FS_{\phi,r}}$ are given 
as in Equations~\eqref{eq-11} and \eqref{eq-12}. Hence, item 2 follows.

\subsection{Proof of Proposition~\ref{prop:critical rate}}

We recall the desingularised system, Equation~\eqref{eq-10}, in which the directionality of the flow coincides with that of the reduced flow on the attracting sheet $S_{0, \phi}^{1,a}$ due to $-(1+a\phi-2\phi u)>0$, while it is reversed on the repelling sheet $S_{0, \phi}^{1,r}$ where $-(1+a\phi-2\phi u)<0$. Furthermore, we recall that the fold curve $F_{\phi}$ is defined by $u_F=\tfrac{1+a\phi}{2\phi}$, and we note that the initial condition $I$ lies on $S_{0, \phi}^{1, a}$, as $u_F<1$ for $\phi\in[\phi_{\min,\delta},\phi_{\max,\delta})$.
By Proposition~\ref{prop-4}, $FS_{\phi, r}$ is a hyperbolic saddle, the stable manifold $W^{\rm s}(FS_{\phi, r})$ of which forms a separatrix for trajectories on $S_{0, \phi}^{1, a}$: as $\dot\phi=r>0$ in Equation~\eqref{eq-9}, the trajectory $\gamma_I$ initiated at $I$ increases in $\phi$; its fate is determined by the position of $I$ relative to $W^{\rm s}(FS_{\phi,r})$. As before, we denote the $\phi$-coordinate of $FS_{\phi,r}$ by $\phi^\ast$.
\begin{enumerate}
\item For $r<r_{\rm crit}$, $I$ lies ``above" $W^{\rm s}(FS_{\phi,r})$, with $\phi>\phi^\ast$. Here, $\dot{u}|_{u=1}=\frac{(1-a)r}{1-(2-a)\phi}<0$, as $\phi\geq \phi_{\rm min, \delta}>\phi_{\rm min}=(2-a)^{-1}$, which implies that $u<1$ on $\gamma_I$ for all $\tau>0$. Next, we recall that $\dot{u}$ is singular at $u=u_{F}(\phi)$; cf.~\eqref{eq-9a}. Therefore, we consider a small perturbation off $F_{\phi}$, with $u=u_F+\rho$ for $\rho>0$ sufficiently small. A simple calculation gives 
\[
\dot{u}|_{u=u_F{(\phi)}+\rho}=-\frac{(1-a\phi)\big\{(1-a\phi)[1-(2-a)\phi]+2r(1+a\phi)\big\}}{16\phi^3}\frac{1}{\rho}+\mathcal{O}(1)\quad\text{as }\rho\to0^+.
\]
The coefficient $\tfrac{1-a\phi}{16\phi^3}$ is positive due to $a,\phi\in(0,1)$, while the term in curly brackets is zero for $\phi=\phi^\ast$ and negative for $\phi>\phi^\ast$; in particular, since $r<r_{\rm crit}$, $\phi_{\rm min,\delta}>\phi^\ast$. Therefore, $\dot{u}|_{u=u_F{(\phi)+\rho}}>0$, which implies that $\gamma_I$ is bounded between $F_{\phi}$ and $\{u=1\}$ for all $\tau>0$ and hence must remain on $S_{0,\phi}^{1,a}$.
\item For $r=r_{\rm crit}$, $I\in W^{\rm s}(FS_{\phi,r})$, i.e., $\gamma_I$ corresponds to $W^{\rm s}(FS_{\phi,r})$ or, equivalently, to the singular canard $\Gamma_0$ itself, flowing to $FS_{\phi,r}\in F_{\phi}$. Since the numerator and denominator in \eqref{eq-9} vanish simultaneously there,
$\gamma_I$ crosses the fold onto the repelling sheet $S_{0,\phi}^{1,r}$.
\item For $r>r_{\rm crit}$, $I$ lies ``below" $W^{\rm s}(FS_{\phi,r})$, with $\phi<\phi^\ast$. Again, we find $\dot{u}|_{u=1}<0$ and, therefore, $u<1$ for all $\tau>0$.
As a result, $\gamma_I$ enters the region bounded by $\{u=1\}$, $F_{\phi}$, and $W^{\rm s}(FS_{\phi,r})$. Due to the invariance of $W^{\rm s}(FS_{\phi,r})$ and $\dot{\phi}=r>0$ for $\phi\in[\phi_{\min,\delta},\phi_{\max,\delta})$, $\gamma_I$ must intersect $F_{\phi}$ in finite time, where it leaves $S_{0, \phi}^{1, a}$ along a fast fibre of the layer flow of \eqref{eq-ramped layer} in the $u$-direction.
\end{enumerate}
Finally, to determine $r_{\rm crit}$, we recall that the stable manifold $W^{\rm s}(FS_{\phi, r})$ is approximated by $\phi=\tfrac{1}{2-a}+\tfrac{2}{(1-a)(2-a)}r+\mathcal{O}(r^2)$, to leading order in $r$. Setting $\phi_{\min,\delta}=\tfrac{1}{2-a}+\tfrac{2}{(1-a)(2-a)}r+\mathcal{O}(r^2)$ and solving to leading order in $\delta$, we find the stated expression for $r_{\rm crit}$.

\subsection{Proof of Corollary~\ref{cor:critical rate}}

The argument follows the proof of Proposition~\ref{prop:critical rate}. Specifically, to calculate the $\mathcal{O}(\delta^2)$-coefficient in the expansion for $r_{\rm crit}$, we need to determine the coordinates of $FS_{\phi, r}$ and the components of the corresponding stable eigenvector $\textbf{v}_-$ to second order. Furthermore, $W^{\rm s}(FS_{\phi, r})$ must be approximated to first order via the invariance equation. We omit the requisite calculations, as the proof is standard, but algebraically involved.

\end{document}